\documentclass[preprint,aps,superscriptaddress,floatfix]{revtex4}
\usepackage{graphicx,epsfig,amssymb,amsmath,array}
\usepackage{relsize,placeins,float}
\usepackage[dvipsnames]{xcolor}
\usepackage{blindtext}
\usepackage[english]{babel}

\begin{document}

\title{When a periodic forcing and a time-delayed nonlinear forcing drive a non-delayed Duffing oscillator. }

\author{Mattia Coccolo}
\affiliation{Nonlinear Dynamics, Chaos and Complex Systems Group, Departamento de F\'{i}sica,
Universidad Rey Juan Carlos, Tulip\'{a}n s/n, 28933 M\'{o}stoles, Madrid, Spain}

\author{Miguel A.F. Sanju\'{a}n}
\affiliation{Nonlinear Dynamics, Chaos and Complex Systems Group, Departamento de F\'{i}sica,
Universidad Rey Juan Carlos, Tulip\'{a}n s/n, 28933 M\'{o}stoles, Madrid, Spain}

\date{\today}
\date{\today}

\begin{abstract}

When two systems are coupled, the driver system can function as an external forcing over the driven or response system. Also, an external forcing can independently perturb the driven system, leading us to examine the interplay between the dynamics induced by the driver system and the external forcing acting on the response system.  The cooperation of the two external perturbations can induce different kinds of behavior and initiate a resonance phenomenon. Here, we analyze and characterize this resonance phenomenon. Moreover, this resonance may coexist in the parameter set and coincide with other resonances typical of coupled systems, as {\it the transmitted resonance} and {\it the coupling-induced resonance}. Thus, we analyze the outcomes to discern their distinctions and understand when the increase in oscillation amplitudes is attributable to one phenomenon, to one of both the others, or a combination of the three.

\end{abstract}

\maketitle

\section{Introduction}

Driven systems play a crucial role in diverse scientific disciplines, ranging from medicine \cite{jiruska, mormann, Rulkov} and physics \cite{jensen, hramov} to communication \cite{Koronovskii, Naderi}, mechanics \cite{defoort}, networks \cite{delellis, zhang}, circuits \cite{Yao}, and engineering \cite{sujith}. They provide insights into phenomena like phase transitions in ferromagnetic materials, investigated through mean-field coupling \cite{Desai, Dawson}. Winfree \cite{Winfree} proposed a model for collective synchronization, representing populations of coupled nonlinear oscillators with globally stable limit cycles. In the realm of neuroscience, driven systems established a qualitative link between the synchronous flashing of fireflies and the origin of brain rhythms \cite{Wiener}. Beyond synchronization, we explore scenarios where these dynamical systems undergo enhancement due to the emergence of resonance.

The study of resonance phenomena holds significant importance, as evident in a rich body of literature. Various forms of nonlinear resonance, such as stochastic resonance \cite{Gammaitoni, McDonnell}, chaotic resonance \cite{Zambrano}, vibrational resonance \cite{Landa}, delay-induced resonance \cite{Cantisan, Coccolo}, and Bogdanov-Takens resonance \cite{Coccolo2}, have been extensively investigated, each named after its specific inducing mechanism. Resonance plays a pivotal role in coupled systems, enhancing overall performance and efficiency, whether it be the coupling of strings in musical instruments, planets in astrophysics, or components in engineering applications.

In a recent study \cite{Coccolo_synch}, the authors employ the continuous control coupling mechanism \cite{Ding, Kapitaniak} to couple two systems through a  parameter of the response system, the coupling constant. They explored the impact of a time-delayed driver system as the sole external force on the oscillation amplitudes of a non-delayed driven system, revealing the emergence of resonance induced by the coupling mechanism  (the coupling-induced resonance). Additionally, in another work \cite{Coccolo_synch2}, the driver system was perturbed with a periodic force, resulting in resonance, which was then transmitted to the response system through the coupling mechanism, this phenomenon has been called {\it transmitted resonance}. Subsequently, we delve into examining the effects on the response system's dynamics when two external forces act on it: the time-delayed driver system and a periodic force. This interaction can lead to abrupt changes in the oscillator behavior and, specifically, giving rise to a resonance phenomenon. Various coupling methods can be used \cite{Pecora1, Pecora2, Pecora3}, but we believe the continuous control is the most direct and provides a straightforward model of the driver system's effects on the response system as an external force.

In summary, our approach involves utilizing both the driver system and a periodic signal as forces for the response system, investigating the conditions under which a distinct resonance phenomenon emerges in the response system's dynamics. In essence, we analyze the interaction between the coupling mechanism and a periodic force driving the response system, exploring situations where this resonance phenomenon coexists and overlaps with the transmitted resonance and the coupling-induced resonance to discern their differences.

 Therefore, after introducing briefly the model in Sec.~\ref{Sec:II},  we analyze the resonance phenomenon in Sec.~\ref{Sec:III}. In particular, we examine the induced resonance in the response system oscillations due to the cooperation between the coupling constant and the external forcing of the response system. Next, we explore how changes in the coupling constant and the amplitude of external forcing in the response system impact its oscillations. Finally, concluding remarks are presented in Sec.~\ref{Sec:concl}.

\section{Model}\label{Sec:II}
We study the response of a Duffing oscillator (the response system) when it is driven by a time-delayed Duffing oscillator (the driver system).  The two systems coupling mechanism and the behaviors of the driver system has been thoroughly studied in \cite{Coccolo_synch, Coccolo_synch2}. Therefore, we just recollect here the most important details. The coupling mechanism is the unidirectional method called the {\it continuous control}  \cite{Ding,Kapitaniak}. This coupling mechanism uses the coupling constant $C$ that multiply the difference of the the positions $(x_1-x_2)$.  The numerical value of the coupling constant  measures the strength of the coupling for the forcing induced by the time-delayed Duffing oscillator. Moreover, an external periodic forcing shows up in the equations of the systems yielding 
\begin{align}\label{eq:1}
Driver&\rightarrow \quad  \frac{d^2x_1}{dt^2}+\mu\frac{dx_1}{dt}+\gamma x_1(t-\tau)+\alpha x_1(1-x_1^2)=0,\\
Response&\rightarrow \quad   \frac{d^2x_2}{dt^2}+\mu\frac{dx_2}{dt}+\alpha x_2(1-x_2^2)=C(x_1-x_2)+f\cos{\omega_2 t}, 
\end{align}
where we fixed the parameters $\mu=0.01$, $\alpha=-1$ and $\gamma=-0.5$. In order to appreciate the effects of the two perturbations on the dynamics of the response system, the dissipation is negligible, $\mu=0.01$. The history functions of the driver system are $u_0=v_0=1$, and the initial conditions of the response system are $x_0=y_0=0.5$. All simulations were performed using Matlab's {\it ddesd} function with adaptive step size. The initial transients have been excluded from the analysis, as they have a negligible impact on the results after the system reaches the steady-state, which is the primary focus of this study. In this scenario the driver system acts as a forcing on the response system through the coupling mechanism, while the periodic forcing is the response system external forcing.

\begin{figure}[htbp]
  \centering
   \includegraphics[width=16.0cm,clip=true]{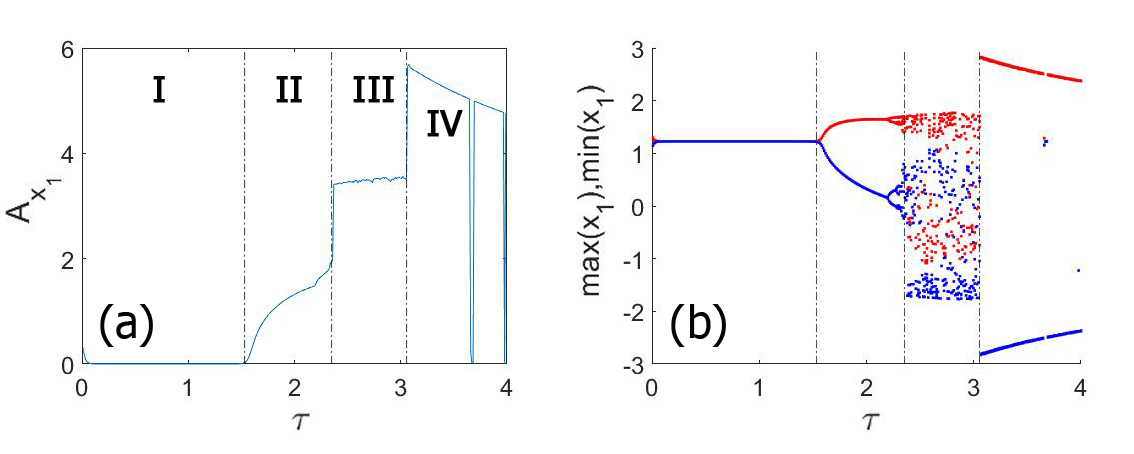}
   \caption{The panels show the oscillation amplitudes ($A_{x_1}$) (a) and the maximum and minimum diagram (b) of the driver system. It is possible to appreciate the oscillation amplitudes (a) and the oscillator behaviors (b) in all the $\tau$ regions of the driver system. The history functions for the driver system are constant $(u_{0},v_{0})=(1,1)$.}
\label{fig:2}
\end{figure}

As it has already been reported in \cite{Coccolo, Cantisan}, the time-delayed Duffing oscillator  without forcing undergoes various bifurcations while $\tau$ changes, describing 4 different regions, as shown in Fig~\ref{fig:2}. 
In the first region for $\tau<1.53$ the oscillations fall on one of the fixed points.  Next, in the second region $1.53<\tau<2.35$ the oscillations are sustained but confined to one of the wells. Therefore, it is feasible to observe the emergence of a resonance in these two regions. Our study shows that the analysis of the other two  regions does not contribute to the goal of this paper.\\ 
Throughout the article, we denote the variable of the driver system as $x_1$ and the variable of the response system as $x_2$.  Moreover, in the next section we fix $C=1.66$, as in \cite{Coccolo_synch,Coccolo_synch2}. That coupling value has been proved that gives an enhancement in the response system oscillation amplitudes even without external forcing. This assures us that any further enhancement is due to the new phenomenon and not as a by-product of the cooperation between the coupling constant and the dynamics of the system. 

%  \begin{figure}[htbp]
%   \centering
%    \includegraphics[width=12.0cm,clip=true]{Fig_1}
%    \caption{The double-well potentials of the Duffing oscillator and the time-delayed Duffing oscillator indicating the stable fixed points. The blue one ans the blue vertical lines are referred to the time-delayed Duffing oscillator, the red one and the vertical red lines to the Duffing without delay.}
% \label{fig:1}
% \end{figure}

\begin{figure}[!htbp]
  \centering
   \includegraphics[width=10.0cm,clip=true]{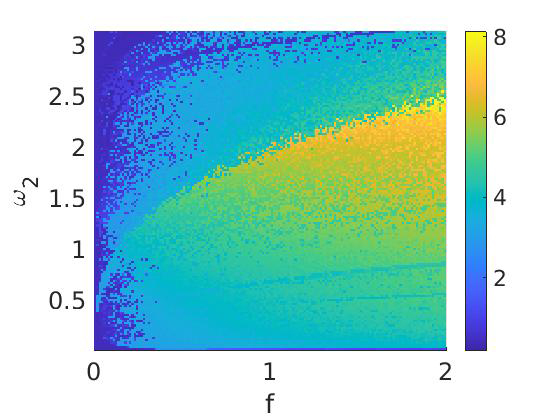}
   \caption{We show the oscillation amplitudes of the response system when the coupling constant $C=0$ in the parameter set $f-\omega_2$.  The figure is meant to be compared with Fig.~\ref{fig:6}(b) and Fig.~\ref{fig:6}(d). The purpose of the comparison is to analyze the effect of the coupling constant on the oscillation amplitudes of the system.}
   \label{fig:3}
\end{figure}

\begin{figure}[htbp]
  \centering
   \includegraphics[width=16.0cm,clip=true]{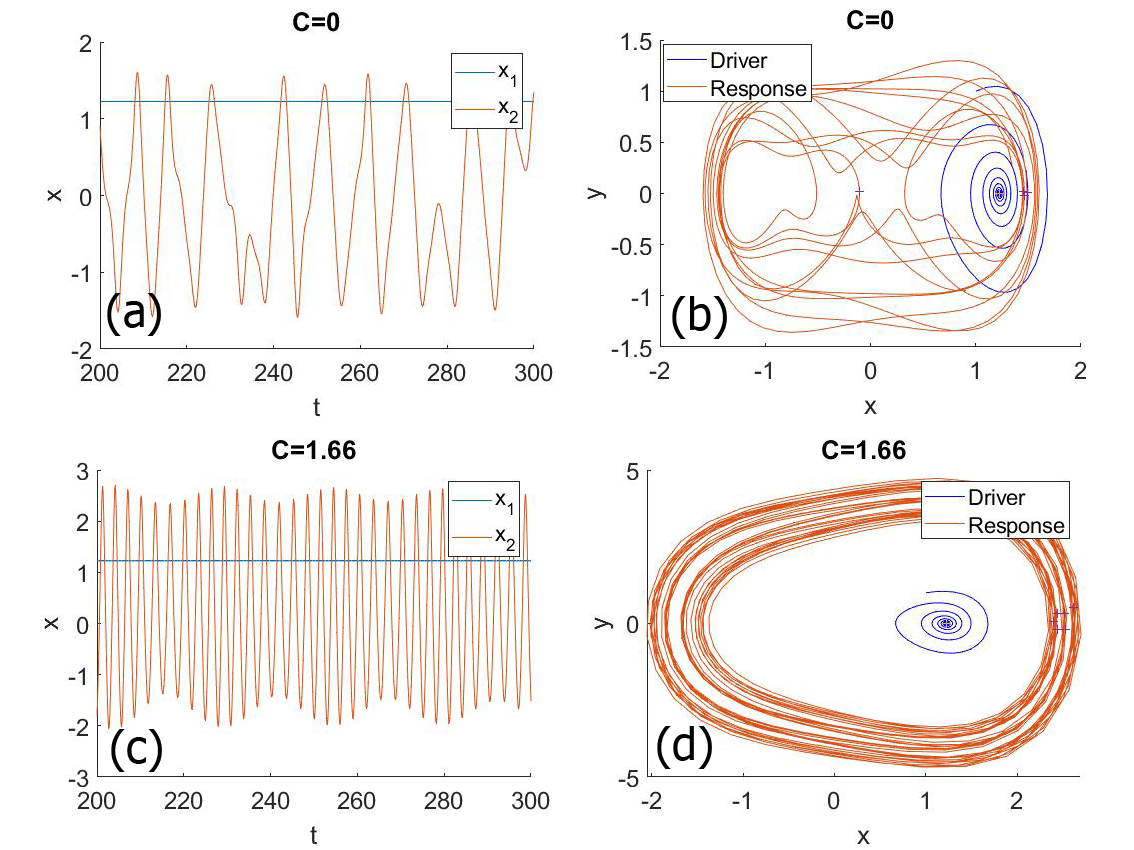}
   \caption{Panels (a) and (b) show the oscillation amplitudes and the trajectories in the phase space of both systems for $C=0$. Panels (c) and (d) show the oscillation amplitudes and the trajectories in the phase space of both systems for $C=1.66$. The other parameters are $F=0,f=0.5,\omega_2=2, \tau=1$.  }
\label{fig:4}
\end{figure}

\begin{figure}[htbp]
  \centering
   \includegraphics[width=16.0cm,clip=true]{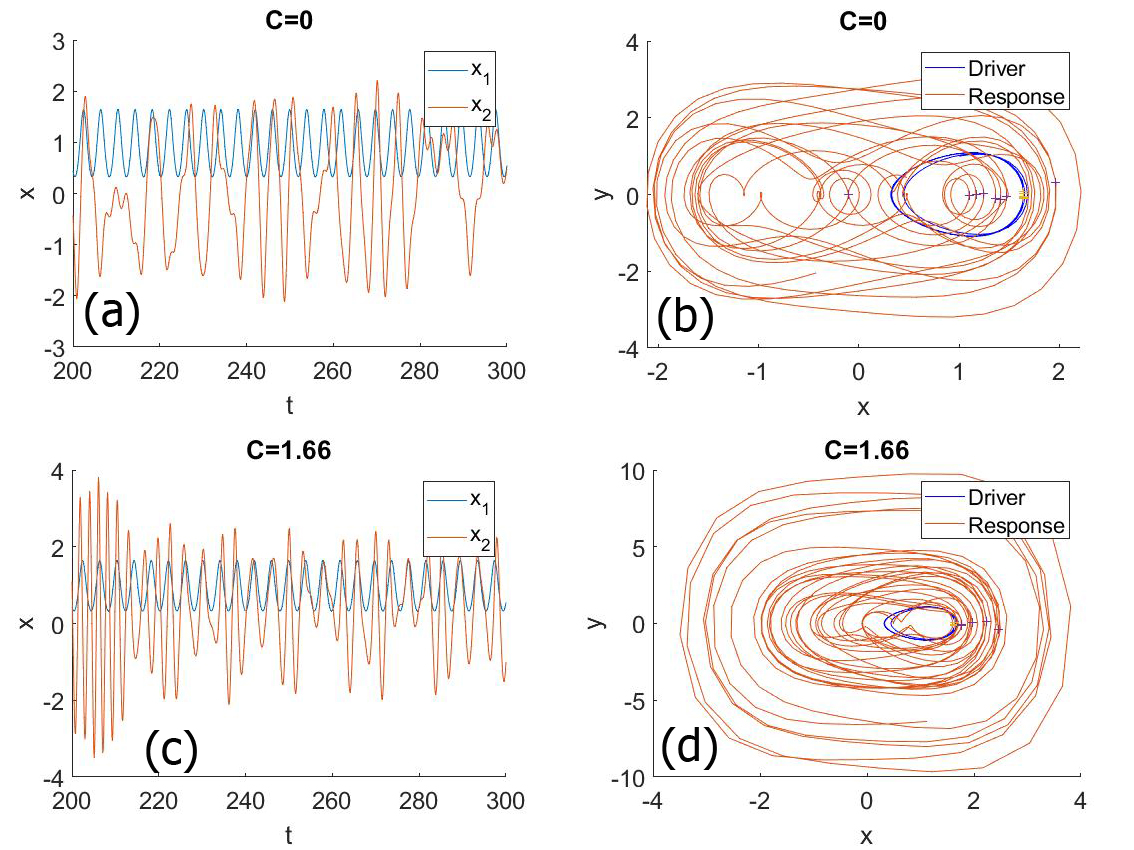}
   \caption{Panels (a) and (b) show the oscillation amplitudes and the trajectories in the phase space of both systems for $C=0$. Panels (c) and (d) show the oscillation amplitudes and the trajectories in the phase space of both systems for $C=1.66$. The other parameters are $F=0,f=1.5,\omega_2=2.5, \tau=2$. }
\label{fig:5}
\end{figure}

\section{coupling-forcing resonance}\label{Sec:III}

Now, we analyze the phenomenon of the resonance induced by the interaction of the coupling term and the external forcing of the response system. First of all, we plot Fig.~\ref{fig:3} to show the amplitude of the response system when $C=0$, to assure that the high oscillation amplitudes analyzed are not typical of the response system but a product of the interaction between the external forcing and the coupling mechanism. Then, in Fig.~\ref{fig:4} we show that high oscillation amplitudes can arise and stabilize thanks to the interaction between the driver system and the external forcing of the response system. In fact, the high amplitudes shown in Fig.~\ref{fig:4}(c) and in Fig.~\ref{fig:4}(d) do not originate from the driver, as illustrated in Fig.~\ref{fig:4}(a) and in Fig.~\ref{fig:4}(b). They are not typical of the response system either, see Fig.~\ref{fig:4}(a) and Fig.~\ref{fig:4}(c). Moreover, in Fig.~\ref{fig:3}, we show the oscillation amplitudes of the Duffing oscillator without coupling for $\omega_2=0.5$, corroborating the oscillation amplitudes shown in Fig.~\ref{fig:4}(a) and in Fig.~\ref{fig:4}(c).

\begin{figure}[htbp]
  \centering
   \includegraphics[width=14.5cm,clip=true]{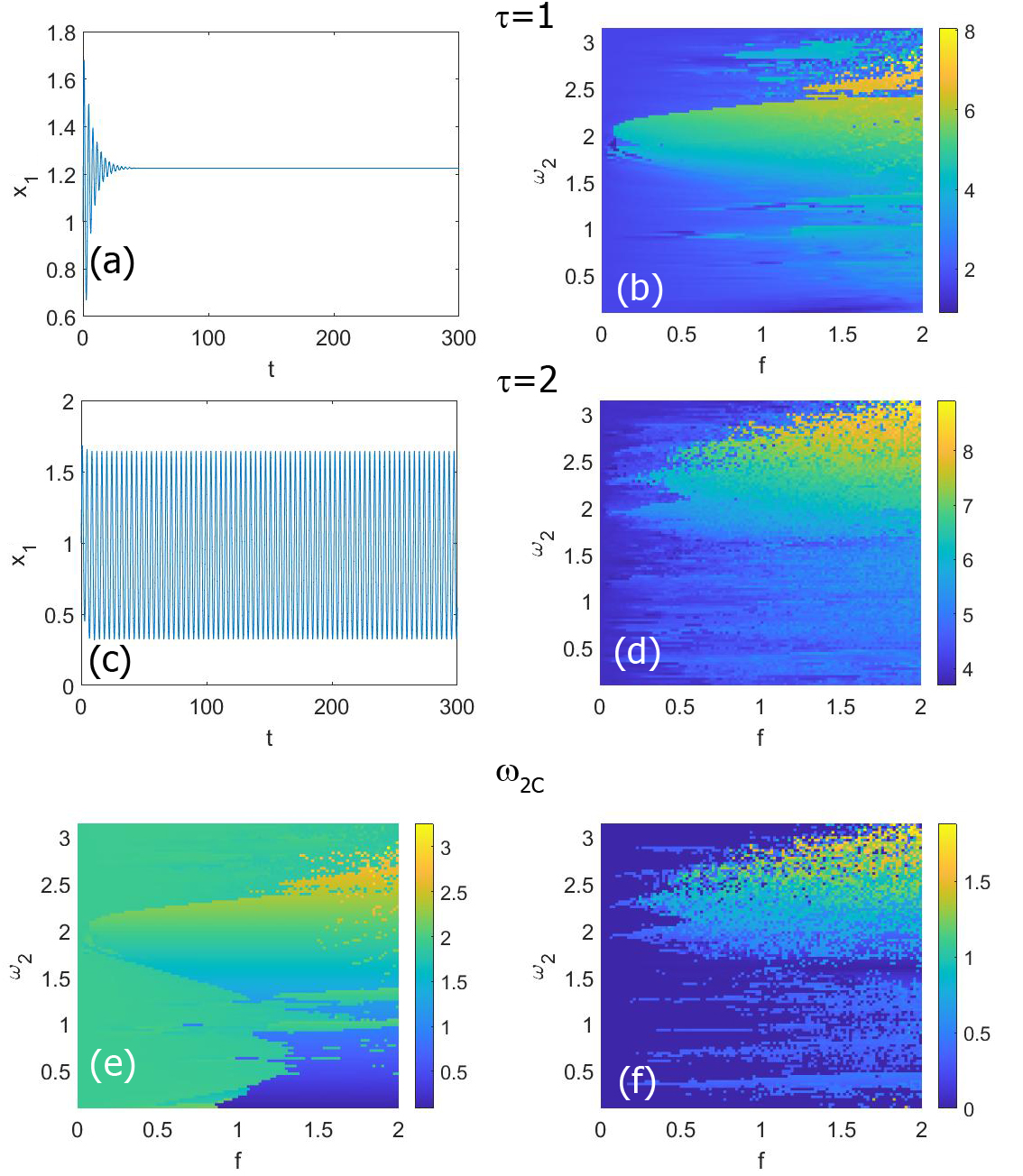}
   \caption{Panels (a) and (c) show the driver system oscillations. Panels (b) and (d) show the oscillation amplitudes of the response system. Panels (e) and (f) show the oscillation frequencies of the response system $\omega_{2C}$, calculated with the FFT, for the two cases above. In all the figures simulations we have set the coupling constant $C=1.66$.  We can appreciate that the driver system does not show high oscillation amplitudes in its asymptotic behavior. On the other hand,  the gradient plot of the oscillation amplitudes in the response system depicts areas of high oscillation amplitudes corresponding to specific values of its external forcing frequency and when the external forcing attains a sufficiently high amplitude. }
\label{fig:6}
\end{figure}

 Figure~\ref{fig:3} is meant to be directly compared with Fig.~\ref{fig:6}(b) and Fig.~\ref{fig:6}(d). In the comparison, one can observe distinctions in the regions exhibiting high oscillation amplitudes between the scenario without the coupling constant and the scenario with a coupling constant $C\neq0$. This contrast serves to  prove that the above mentioned high oscillation amplitudes are not inherent to the system but are instead induced by the driver system, which acts as a forcing, interacting with the external forcing of the response system. We call this phenomenon {\it coupling-forcing resonance}. 
With our understanding of the phenomenon validated, we can now delve into its analysis.

\begin{figure}[htbp]
  \centering
   \includegraphics[width=16.0cm,clip=true]{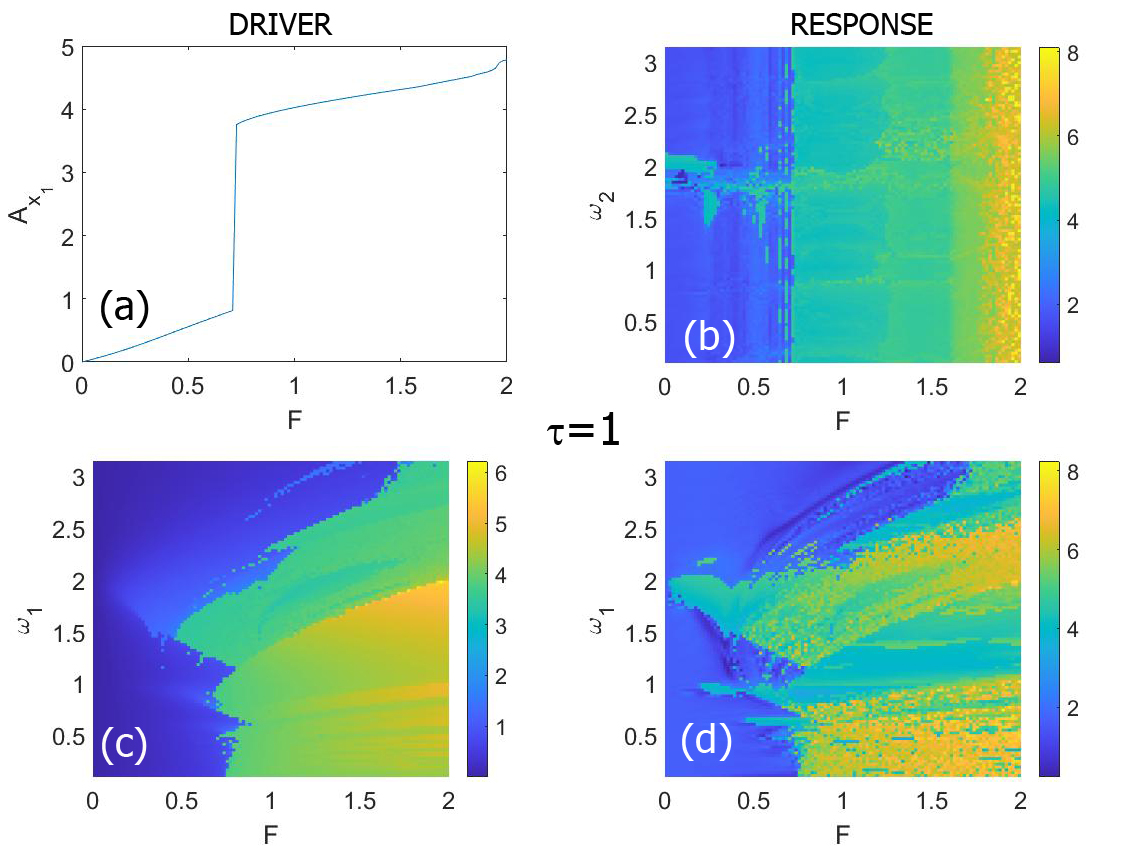}
   \caption{Panel (a) show the driver system oscillations and panel (b) the response system oscillation amplitudes for $f=0.1$ in the parameter set $F-\omega_2$. Panel (c) show the oscillation amplitudes of the driver system and panel (d) of the response system for $f=0$, both in the parameter set $F-\omega_1$. In panel (b)  we can spot a large region of transmitted resonance for values of $F>0.7$ that span all the $\omega_2$ values, and for small values of $F$ around $\omega_2\simeq 2$ a region of resonance induced by the interaction of the coupling constant and the response system external forcing. In panel (d) we can recognize a region of transmitted resonance, and a region of resonance induced by the coupling constant, for small values of $F$ and $\omega_2\simeq 2$. }
\label{fig:7}
\end{figure}

\subsection{The effect of different $\tau$ values}\label{ss:Cf}

 Now, to better understand the phenomenon, we keep on studying other cases of the coupling-forcing resonance, see Fig.~\ref{fig:6}. 
As before, we compare the high oscillation amplitudes shown in this last figure with the oscillation amplitudes of the Duffing oscillator without coupling constant, see Fig.~\ref{fig:3}. Here, we show the gradient plot in the parameter set $f-\omega_2$ of the oscillation amplitudes of the response system without coupling. The comparison with Figs.~\ref{fig:6}(b) and~\ref{fig:6}(d) shows that the high oscillation amplitudes in these two figures are not typical of the Duffing oscillator and Figs.~\ref{fig:6}(a) and~\ref{fig:6}(c) assure that they are not transmitted from the driver to the response system. Again, the explanation of those high oscillation amplitudes could only come from the interaction between the driver system, through the coupling constant, and the response system external forcing. Again, the phenomenon and its explanation is clear and we can keep on with our analysis.

In Fig.~\ref{fig:6}, despite the $\tau$ values belonging to distinct regions, the plots share common factors. One factor is the emergence of resonance around $\omega_2\approx2$. Thus, when the $\tau$ values are within regions I and II, it indicates that the response system is prone to exhibit the coupling-forcing resonance for the external forcing frequency $\omega_{2}=2$. Another common factor is shown in Figs.~\ref{fig:6}(a) and~\ref{fig:6}(c), where the driver does not exhibit any sign of resonance, while the response system does. Subsequently, a distinction becomes evident: in Figs.~\ref{fig:6}(b) and~\ref{fig:6}(d), one can observe that the mist of high oscillation amplitudes is more distinct in the latter figure than in the former. Specifically, in the first figure, the dynamics of the driver system converges to a fixed point, whereas in the second one, the dynamics leads to a limit cycle. We posit that this observation is connected to the diffusion of the high oscillation amplitudes point in the plot. Finally, Figs.~\ref{fig:6}(d) and~\ref{fig:6}(f) show the oscillation frequencies of the response system $\omega_{2C}$ calculated with the FFT. These last panels show that the areas of the coupling-forcing resonance are recognizable also in the frequencies domain.

Then, when two external forcing affect both the driver and response systems obtaining the system equations

\begin{align}\label{eq:2}
&  \frac{d^2x_1}{dt^2}+\mu\frac{dx_1}{dt}+\gamma x_1(t-\tau)+\alpha x_1(1-x_1^2)=F\cos{\omega_1 t}\\
&  \frac{d^2x_2}{dt^2}+\mu\frac{dx_2}{dt}+\alpha x_2(1-x_2^2)=C(x_1-x_2)+f\cos{\omega_2 t}.
\end{align}
The coexistence of this resonance phenomenon with the transmitted resonance,  that has been studied in \cite{Coccolo_synch2}, is confirmed in Figs.~\ref{fig:7}(a) and~\ref{fig:7}(b). The transmitted resonance is a phenomenon for which a resonance triggered in the driver system by its own forcing is transmitted into the dynamics of the response system through the coupling mechanism. In fact,  by comparing Figs.~\ref{fig:7}(a) and~\ref{fig:7}(b)  we can see, in the parameter set $F-\omega_2$, a small zone of higher oscillation amplitudes for small $0.7\lesssim F$ and around $\omega_2=2$. The higher oscillation amplitudes in that zone are not transmitted from the driver system, because they do not appear in Fig.~\ref{fig:7}(a). Moreover, the mentioned zone continues for higher $F$ values although it is submerged inside high oscillation amplitudes result of the transmitted resonance. Those high oscillation amplitudes are transmitted because they find a counterpart in the oscillation amplitudes of the driver in Fig.~\ref{fig:7}(a). Also. the transmitted resonance can coexist with the resonance induced by the coupling constants. In the last Figs.~\ref{fig:7}(c) and~\ref{fig:7}(d), we change the parameter set to $F-\omega_1$ with $f=0$ and the comparison of the two figures shows the coupling forcing induced resonance \cite{Coccolo_synch} around $\omega_1=2$ until the high oscillation amplitudes of the driver kick in and the transmitted resonance starts. Moreover, the three phenomena can be found mixed in the same parameter set, but it can be difficult to discriminate them. In fact, especially for $\tau=2$, some mixed cases have been found but they were not as clear as the ones proposed in the figures above. In fact, the areas that show the three  resonance phenomena are too mixed among each other to be distinguished. This scenario undermines our ability to analyze the phenomena thoroughly.

\subsection{The coupling constant effect} \label{ss:CCr}

\begin{figure}[htbp]
  \centering
   \includegraphics[width=16.0cm,clip=true]{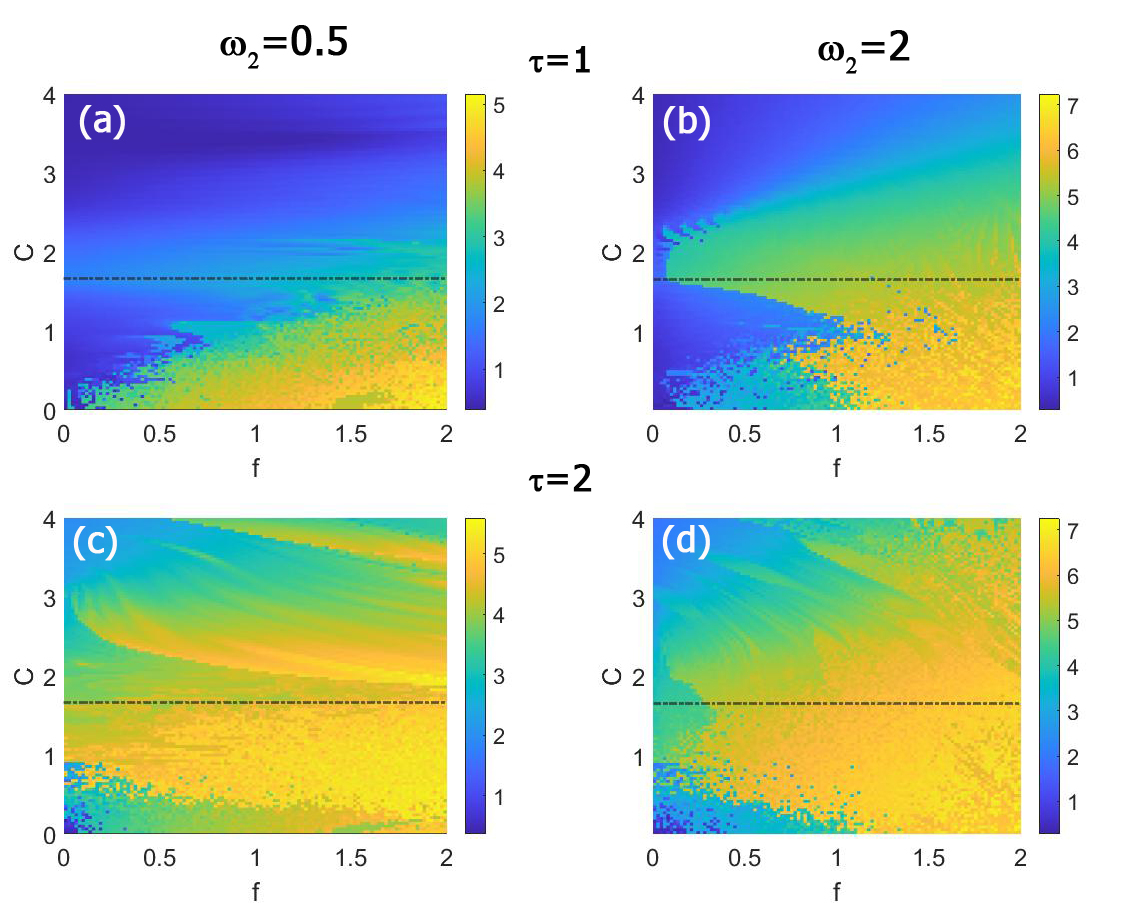}
   \caption{We have set the driver system external forcing amplitude $F=0$. In all the panels we show the oscillation amplitudes gradient in function of the coupling constant $C$ and the response system external forcing amplitude $f$. In the left panels $\omega_2=0.5$ and in the right panels $\omega_2=2$. The latter has previously been demonstrated as the value of the external forcing for the response system, triggering resonance. The black line indicates the oscillation amplitudes at the $C=1.66$ value.}
\label{fig:8}
\end{figure}

Finally, we comment on the ubiquity of the external forcing frequency value $\omega_{1,2}=2$, that seem to play a decisive role for the coupling-forcing resonance, as it also does for the transmitted resonance. In this last phenomenon, for $\tau=1$, this value is the frequency of the driver external forcing to show the resonance that is transmitted to the response system. In the other case, it is the response system external forcing frequency at which the coupling constant and the forcing resonate spontaneously. In fact, not only there is no sign of high oscillations in the two system separately, but also the higher oscillations due to the cooperation of the two external forcing are localized around $\omega_2\simeq 2$. Therefore, it could be worth to study thoroughly the role of this frequency value in function of the coupling constant and the response system external forcing amplitude, in order to compare the results with another value, for example $\omega_2=0.5$.
In Fig.~\ref{fig:8}, we show the effect of the coupling constant on the coupling-forcing resonance in the different $\tau$ regions and for the two aforementioned different frequencies of the response system forcing, $\omega_2$. The driver forcing amplitude is fixed at $F=0$, as in the Fig.~\ref{fig:6}, to be sure that we are studying the resonance induced by the coupling and the external forcing interplay phenomenon without resonance transmission. 

In Fig.~\ref{fig:8}, it is possible to appreciate that the oscillation amplitudes in the right column is higher. In fact, the right column is the one related with the frequency at which the coupling-forcing resonance takes place, as seen in the previous subsection.  In Fig.~\ref{fig:8}(b), we can see that the higher oscillation amplitudes are more generalized and more intermingled with the lower oscillation amplitudes than in Fig.~\ref{fig:8}(a). Also, in Fig.~\ref{fig:8}(b)  an interesting high oscillation amplitudes zone in a triangle shape is recognizable with one of its vertices pointing to $C\approx2$. Finally, Figs.~\ref{fig:8}(c) and~\ref{fig:8}(d) show both large areas of high oscillation amplitudes and no recognizable structures or patterns, although in the $\omega_2=2$ case the oscillation amplitudes are higher. Another aspect worth considering is that Fig.~\ref{fig:8} provides justification for utilizing $C=1.66$. Indeed, this particular value of $C$ consistently corresponds to elevated oscillation amplitudes, as evidenced by the presence of the black line across all panels.

Finally, our analysis of the coupling constant's influence on the onset of the resonance, reveals a complex interplay between the coupling constant and the forcing amplitude of the response system, which varies distinctly across the two $\tau$ regions.

\section{Conclusions}\label{Sec:concl}

Our research involves two coupled systems: a time-delayed Duffing oscillator serving as the driver and a Duffing oscillator without delay as the response system. The former acts as the driving force for the latter. Consequently, each system experiences external forcing independently, or both systems are concurrently influenced by two external forcings. The goal of this work is to investigate the interplay between the coupling mechanism and the response system external forcing. We also analyze the resonance induced by the interaction of the coupling constant and the response system external forcing when another external forcing acts on the driver system. We found that the driver system as external forcing produces a resonance when interacting with the response system external forcing at a specific response system forcing frequency $\omega_2=2$. Consequently, substantial oscillations resulting from the resonance phenomenon arise from this interaction, which are absent in the oscillations of both the driver system and the response system individually. This indicates that the presence of both external perturbations is necessary to induce them. We term this phenomenon the coupling-forcing resonance. Subsequently, we demonstrate that this resonance phenomenon can coexist in a mixed and overlapped manner with the transmitted resonance and the coupling-induced resonance phenomena within the same parameter set. We have, also, investigated how altering the coupling constant and the frequencies of the response external forcing impacts the coupling-forcing resonance. In the $f-C$ parameter set, we fix the external forcing frequency at two distinct values. The outcome reveals that the oscillation amplitudes are more pronounced in the plots when $\omega_2=2$, the frequency that triggers the resonance. The alternative frequency value, $\omega_2=0.5$, was selected for the purpose of comparison.

To conclude, our findings underscore the intricate dynamics underlying the coupling and the external forcing interactions in coupled oscillatory systems. The emergence of coupling-forcing resonance highlights the critical role of both external perturbations in driving novel oscillatory behavior that the driver system cannot achieve when the two forcings act independently. This phenomenon adds depth to our understanding of how resonance can arise from complex interdependencies between coupling mechanisms and external forcings. Moreover, the coexistence and overlap of this resonance with transmitted and coupling-induced resonances in the parameter space further emphasize the rich and multifaceted nature of these interactions.

This study underscores the importance of tuning coupling constants and forcing frequencies, as evidenced by the enhanced resonance effects at specific frequencies. These insights not only deepen our understanding of resonance phenomena but also pave the way for further investigations into the broader implications of such interactions in nonlinear systems.

\section{Acknowledgment}

This work was supported by the Spanish State Research Agency
(AEI) and the European Regional Development Fund (ERDF, EU)
under Project Nos. PID2019-105554GB-I00 and PID2023-148160NB-I00 (MCIN/AEI/10.13039/
501100011033). It has also been supported by Rey Juan Carlos University, Spain, through the Programa Propio de Fomento y Desarrollo de la Investigación de la URJC under the project NABEH, funded by Financiación para proyectos Impulso (Project No. 2024/SOLCON-137648).

%\FloatBarrier

\end{document}